\documentclass[twocolumn,prb,aps]{revtex4}
\usepackage[dvips]{graphicx}
\usepackage[dvips]{graphics}
\usepackage{ulem}
\usepackage{array}
\usepackage{multirow}
\usepackage{hhline}
\usepackage{bigstrut}
\usepackage{slashbox}
\newlength{\bxwidth}\bxwidth=0.8\textwidth

\begin{document}
\title{Dispersion of  the odd magnetic resonant mode in near-optimally doped ${\bf Bi_2Sr_2CaCu_2O_{8+\delta}}$}

\author{
B.~Fauqu\'e$^1$, Y.~Sidis$^1$, L.~Capogna$^{2,3}$, A.~Ivanov$^3$, K.~Hradil$^4$,
C. Ulrich$^2$, A.I.~Rykov$^5$, B.~Keimer$^2$, and P.~Bourges$^{1 \ast}$}

\affiliation{
$^1$ Laboratoire L\'eon Brillouin, CEA-CNRS, CE-Saclay, 91191 Gif sur Yvette, France.\\
$^2$ Max-Plank-Institute f\"ur Festk\"orperforschung, Heisenbergstr. 1, 70569 Stuttgart, Germany.\\
$^3$ Institut Laue-Langevin, 6 Rue J. Horowitz, 38042 Grenoble cedex 9, France.\\
$^4$ Forschungreaktor M\"unchen II, TU M\"unchen, Lichtenbergstr, 1 95747 Garching, Germany.\\
$^5$ Department of Applied Chemistry, University of Tokyo, Hongo 7-3-1, Bunkyo-ku, Tokyo 113-8656, Japan.
}


\pacs{PACS numbers: 74.25.Ha  74.72.Hs, 25.40.Fq }

\begin{abstract}
{ We report a neutron scattering study of the spin excitation
spectrum in the superconducting state of slightly overdoped $\rm
Bi_2Sr_2CaCu_2O_{8+\delta}$ system
($T_c$=87 K). We focus on the dispersion of the resonance peak in
the superconducting state that is due to a S=1 collective mode.
The measured spin excitation spectrum bears a strong similarity to
the spectrum of the $\rm YBa_2Cu_3O_{6+x}$ system for a similar
doping level ({\it i.e.} $x\sim 0.95-1$), which consists of
intersecting upward- and downward-dispersing branches. A close
comparison of the threshold of the electron-hole spin flip
continuum, deduced from angle resolved photo-emission measurements
in the same system, indicates that the magnetic response in the
superconducting state is confined, in both energy and momentum,
below the gapped Stoner continuum. In contrast to $\rm
YBa_2Cu_3O_{6+x}$, the spin excitation spectrum is broader than
the experimental resolution. In the framework of an
itinerant-electron model, we quantitatively relate this intrinsic
energy width to the superconducting gap distribution observed in
scanning tunnelling microscopy experiments. Our study further
suggests a significant in-plane anisotropy of the magnetic
response.}

\end{abstract}

\maketitle

\section{Introduction}

It is now well established that the spin excitation spectrum in
the superconducting (SC) state of many high-$T_c$ superconductors is
dominated by an unusual  spin triplet excitation
\cite{Sidis_review,rossat91,mook93,fong95,bourges96,fong96,miami,fong00,dai01,Nature_Fong99,he,tl2201}.
This excitation is referred to as the {\it magnetic resonance
peak}. It is centered at the planar antiferromagnetic (AF) wave
vector ${\bf q}_{AF}=(\pi/a,\pi/a)$ (where $a$ is the lattice
spacing) and at an energy $\rm E_r$ that scales with the SC
critical temperature, $T_c$. Further, the resonance peak intensity
vanishes above $T_c$ and exhibits a temperature dependence that
looks like an order parameter. A renormalization of its
characteristic energy with temperature is not observed within the
experimental error \cite{bourges96,fong96}. This behavior has been
reported in several families of copper oxides with maximum SC
critical temperatures $ T_c^{max} \ge $90 K: in ${\rm
Tl_2Ba_2CuO_{6+x }}$ with uniformly spaced, single CuO$_2$ layers
\cite{tl2201}, as well as in bilayer systems such as $\rm
YBa_2Cu_3O_{6+x}$ (YBCO)
\cite{rossat91,mook93,fong95,bourges96,fong96,miami,fong00,dai01}
and $\rm Bi_2Sr_2CaCu_2O_{8+x}$ (Bi2212) \cite{Nature_Fong99,he}.
The magnetic resonance peak indicates the existence of a S=1
collective mode with a peculiar dispersion. Its downward
dispersion starting at {\bf q}$\rm_{AF}$ was first observed in
YBCO \cite{science_Bourges00}. Complementary
measurements provided strong indications of a second upward
dispersion starting from the resonance energy in optimally doped
YBCO\cite{PRL_Pailhes04,PRL_Reznik04,woo}. Importantly, both
dispersive branches vanish above $T_c$ as does the magnetic
resonance peak at ${\bf q}_{AF}$. In strongly underdoped cuprates
($T_c \le 62$ K), similar dispersive excitations have also been
reported \cite{bourges97,stock,hayden04,Nature_Hinkov04,hinkov06},
but only the downward branch vanishes above $T_c$ whereas the
upward dispersion remains essentially unchanged across
$T_c$\cite{hinkov06}. Therefore, the spin excitation dispersion in
YBCO exhibits a "hour glass"-like shape centered at the resonance
peak.

Recently, the debate became focused on the origin of the S=1
dispersive collective mode. The theoretical description of the
mode is important, because antiferromagnetism is generally
believed to play a significant role in the SC pairing mechanism in
high-$T_c$ cuprates \cite{scalapino}.

First, based on the spin dynamics data in the stripe ordered
system $\rm La_{7/8}Ba_{1/8}Cu O_{4}$, it has been proposed that
$E_r$ could be a saddle point in the dispersion, with spin
excitations propagating along a given in-plane direction (say
$a^*$) below $E_r$ and along the perpendicular direction  ({\it
i.e.} $b^*$) above $E_r$ \cite{jmt04,revue_jmt}. This gives rise
to an X-like shape in twinned crystals where $a^*$ and $b^*$ are
mixed. Saddle points can arise in models with spin and charge
stripe order at low temperature. The spin excitation spectrum can
then be modelled by a specific bond-centered stripe model
according to which non-magnetic charge stripes separate a set of
weakly coupled two-leg spin ladders in the copper oxide layers.
The low energy excitations (below $E_r$) correspond to collective
excitations that propagate perpendicular to the ladder direction,
whereas the high energy part of the spectrum (above $E_r$) is
associated to intra-spin ladder excitations propagating parallel
to the lines of charges. This picture, later on sustained by
calculations \cite{Voj04,Uhr04,Sei05}, implies a pronounced
in-plane anisotropy of the magnetic spectrum. However, this is not
consistent with the spin excitations in SC cuprates, in particular
in underdoped and optimally doped YBCO
\cite{Nature_Hinkov04,hinkov06}. Indeed, using detwinned YBCO
samples \cite{hinkov06}, it has been shown that the spin
excitation spectrum exhibits a 2D geometry both below
\cite{Nature_Hinkov04} and above \cite{hinkov06} $E_r$,
inconsistent with a saddle-point dispersion. However, recent
calculations of the spin dynamics considering fluctuating stripe
segments \cite{Voj06}, shows that the magnetic spectrum is losing
its 1D character for short charge segments (a few atomic
distances) because they actually exist in both perpendicular
directions. Further, it should be emphasized, once again, that the
resonance peak intensity in all SC cuprates exhibits
systematically a strong temperature dependence in the SC state
\cite{Sidis_review}. This is in a marked contrast to the data in
the stripe-ordered system, where this anomalous temperature
dependence is absent. These inconsistencies cast some doubt about
a similar origin of the resonance peak seen in cuprates where
superconductivity is well developed \cite{Sidis_review} and the
one reported in the stripe-ordered system \cite{jmt04}. As a
matter of fact, it would be more meaningful to compare the
spectrum in non-SC $\rm La_{7/8}Ba_{1/8}Cu O_{4}$ with the
magnetic spectrum in the normal state of other SC cuprates, as it
has recently been done in underdoped YBCO \cite{hinkov06}.

Second, starting from the metallic side of the phase diagram of
high-$T_c$ superconductors and reducing the hole doping, one can
try to understand the  S=1 collective mode within an
itinerant-electron model. It has been proposed that the resonant
magnetic collective mode could be described as {\it a spin
exciton}
\cite{Eschrig_review,PRL_Chubukov99,norman,PRB_Onufrieva02,PRB_Schnyder04,PRL_Eremin05},
i.e a S=1 bound state, pushed below the gapped Stoner continuum in
SC state by AF interactions. This type of excitation exists in the
SC state only and vanishes when the gap disappears in the normal
state. Alternatively, when starting from the Mott-insulator side
of the phase diagram and increasing the hole doping, in framework
of localized-spin models the mode can be viewed as the remnant of
the magnon observed in the insulating AF state
\cite{PRB_Sega03,condmat_Prelovsek06}.
The collective modes of localized spins on Cu sites may survive in
the metallic state, but are heavily damped by scattering from
charge carriers. Long-lifetime collective excitations can then be
restored in the SC state, when scattering processes are eliminated
below the gapped Stoner continuum. When the mode energy is close
to the gap, it can be viewed as a spin-exciton, as in the
itinerant-spin approach \cite {PRB_Sega03,condmat_Prelovsek06}.
Both approaches represent two different limits of a dual
description of the magnetism of high-$T_c$ superconductors:
localized and itinerant spins are tightly bound and cannot be
disentangled \cite{PRB_Onufrieva94,JETP_Eremin06}. It is worth
emphasizing that in a dual approach one can schematically ascribe
the upper dispersion to the localized character of the magnetic
response and the lower one to the itinerant one. However, 
future quantitative calculations are necessary to validate this picture. 

In all of these models, the change of the band electronic
excitations upon passing through $T_c$ has an important feedback
on the spin excitation spectrum in the SC state. The determination
of the electron-hole spin flip continuum then requires a good
knowledge of the fermionic dispersion relations, which is still
in a stage of rapid development in the YBCO system
\cite{condmat_Borisenko06}. In contrast, the charge excitations in
$\rm Bi_2Sr_2CaCu_2O_{8+\delta}$ have been extensively studied by
surface-sensitive techniques. First, the Fermi surface and the
band dispersions have been determined by angle resolved
photo-emission spectroscopy (ARPES) (see
\cite{Eschrig_review,arpes,PRB_Kordyuk03} and references therein).
Second, Scanning Tunnelling Microscopy (STM) data
\cite{PRB_Howald,PRL_McElroy05,Nature_Davis06} evidence a local
distribution of the superconducting gap and then infer the low
energy quasi-particle excitations and the Fermi surface through
Friedel oscillations. Recently, attempts to compute the spin
excitation spectrum in nearly optimally doped Bi2212 starting from
ARPES measurements \cite{condmat_Campuzzano06,borisenko} have been
performed using the itinerant-spin approach (excitonic scenario).

In this paper, we present a study of both energy and momentum
dependences of the resonant spin excitations in the SC state of a
nearly optimally doped Bi2212.  Our study reveals that the spin
excitation spectrum in Bi2212 bears close similarity to the spin
excitation spectrum reported in YBCO for the same doping level. We
further show that the resonant spin collective modes are located
below the gapped Stoner continuum computed from ARPES data. The
combination of our inelastic neutron  scattering (INS)
measurements and ARPES and STM measurements on the same system
provides the opportunity to test theoretical models for the spin
dynamics in the cuprates.

\section{Experimental}

For the present study, we used a single crystal of of slightly
overdoped Bi2212 with volume $\sim$1.5 g and $T_c$=87 K . The INS
measurements were performed on the thermal triple-axis
spectrometers IN8 at the Institute Laue Langevin in Grenoble
(France), 2T at the Laboratoire L\'eon Brillouin at the reactor
Orph\'ee in Saclay (France), and PUMA at the reactor FRM-II in
Garching (Germany). Measurements were carried out with a
double-focusing PG(002) monochromator and a PG(002) analyzer. The
final neutron wave vector was set to $k_f$=4.1~\AA$^{-1}$,
yielding an energy resolution $\sigma_{\omega}\simeq$6 meV. A PG
filter was inserted into the scattered beam in order to eliminate
higher order contaminations. For high energy transfers  ($\hbar
\omega >$55 meV), the PG filter was removed and $k_f$ set to
5.5~\AA$^{-1}$. To cover the full q-dependence of the spin
excitations, three different scattering planes were used for the
measurements. The sample was successively oriented such that
momentum transfers $\bf Q$ of the form (H,H,L), (K/3,K,L) and
(H,0.3L,L) were accessible. We use a notation in which $\bf Q$ is
indexed in units of the tetragonal reciprocal lattice vectors
$2\pi/a=1.64$\AA$^{-1}\equiv 2\pi/b$ and $2\pi/c=0.203
$\AA$^{-1}$.

\begin{figure}
\centerline{\includegraphics[angle=0,width=8.5 cm, height=10 cm]{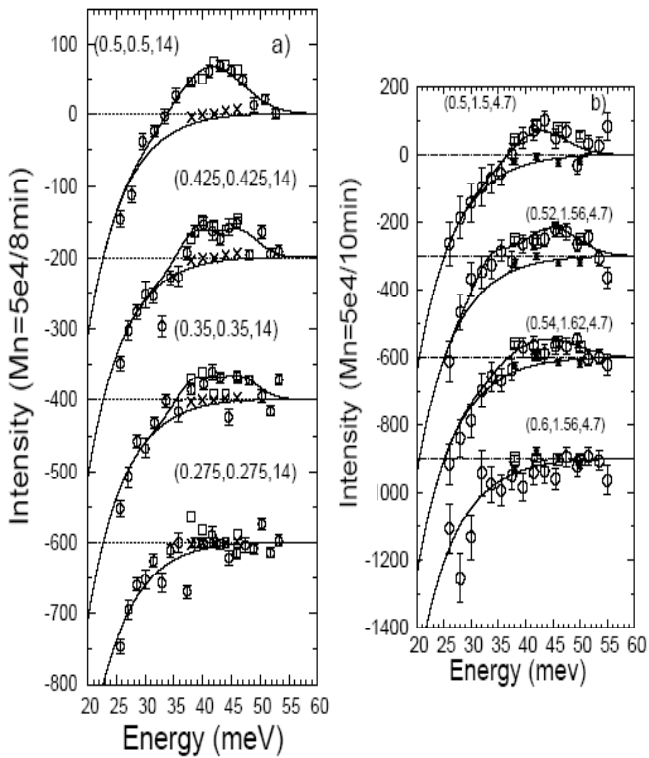}}
\caption{ Difference between constant-Q scans performed at 10 K and 100 K. INS measurements
were carried out on spectrometer 2T. The differential spectra are shown at different
wave vectors of the form: a) {\bf Q}=(H,H,L=14), b) {\bf Q}=(K/3,K,L=4.2). In addition
to the  energy scan data (open circles) are also reported the magnetic intensities
(open squares) and the location of the negative background (crosses), deduced from
the analysis of  constant energy scans (Fig.~\ref{Fig-Qscan}). Those data , obtained
on spectrometer IN8, have been rescaled. The lines indicate the energy dependence
of the negative background (see text) and the enhancement of the magnetic intensity is
described by a set of Gaussian functions on top of the negative background.}
\label{Fig-Escan}
\end{figure}

The magnetic neutron scattering cross section is proportional to
$Im \chi ({\bf Q},\omega)$, the imaginary part of the dynamical
magnetic susceptibility, weighted by the square of the Cu magnetic
form factor, $F(Q)$, and the detailed balance temperature factor
\cite{bourges96,fong96,fong00}. For a paramagnetic system, it
reads:
\begin{equation}
{{d^2\sigma}\over{d\Omega d\omega}} = {{r_0^2}\over{2\pi}}  |F(Q)|^2
{1\over{1-\exp(-{{\hbar\omega}\over{k_B T}})}}  Im \chi ({\bf Q},\omega)
\label{sab}
\end{equation}
where ${\bf Q}=(H,K,L)$ is the full wave vector and  ${\bf
q}=(H,K)$ is the planar wave vector in the CuO$_2$ plane. $r_0
=0.54\ 10^{-12}$ cm is the neutron magnetic scattering length.
Like the YBCO system, the Bi2212 system contains two $\rm CuO_2$
planes per unit cell. Owing to the interaction between the $\rm
CuO_2$ planes within a bilayer, spin excitations with odd (o) and
even (e) symmetry with respect to the exchange of the layer
contribute to the spin susceptibility as in YBCO
\cite{fong00,PRL_Pailhes03}, so that:
\begin{equation}
\chi({\bf Q},\omega) =  \sin^2 (\pi z L) \chi_{o}({\bf q},\omega) + \cos^2(\pi z L)\chi_{e}({\bf q},\omega)
\label{eq-bilayer}
\end{equation}
$z$=0.109 is the reduced distance between the $\rm CuO_2$ planes
of the bilayer. Using different $L$ values of the c$^*$ component
of the momentum transfer, the mode of each symmetry can be
measured. The observation of resonant modes with both symmetries
has recently been reported in Bi2212 \cite{Lucia} in two samples
and in particular in the sample studied here. In the present
study, we focus on the odd spin excitations, that can be
selectively probed by an appropriate choice of the L component:
L=14  for the planar wave vector (0.5,0.5) and L=4.7 for the
planar wave vector (0.5,1.5). Note that the square of the Cu
magnetic form factor is about 1.4 times larger for (0.5,1.5,4.7)
than for (0.5,0.5,14). This is because the magnetic form factor is
anisotropic in the cuprates \cite{shamoto}.

\begin{figure}
\centerline{\includegraphics[angle=0,width=8.5 cm, height=10 cm]{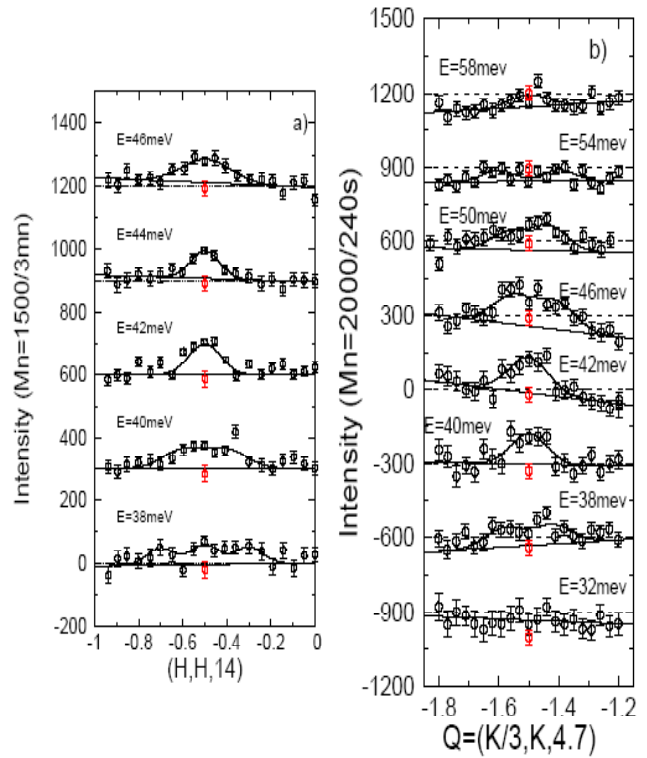}}
\caption{Differences between constant energy scans performed at 10 K and 100K:
a) scans along the (110) direction, b) scans along the (130) direction.
All measurements were carried out on the spectrometer IN8. The position of the
negative background used in the analysis of the energy scans, reported in
Fig~\ref{Fig-Escan}, is also shown at the AF wave vector (red circles). The solids
lines correspond to the fit of the data by a single or double Gaussian functional
form, on top of a sloping background.}
\label{Fig-Qscan}
\end{figure}

\section{Inelastic Neutron Scattering measurements}


Throughout this paper, we focus on the enhancement of the magnetic
response in the SC state, corresponding to the resonant spin
excitations. To extract this response, we subtracted scans at T=10
K and T=100 K ($> T_c$=87 K). This procedure generates a negative
background in the differential spectra, owing to the thermal
enhancement of the nuclear background. The main contribution to
this negative background comes from the thermal population of
phonons given by the detailed balance factor. In the energy range
of interest for the present study, 25-60 meV, the nuclear
(phononic) background continuously decreases. Thus, the energy
dependent negative background can be quite well approximated by a
functional form like $(a+b\omega)/(\exp(-\frac{\hbar \omega }{k_B
T_o})-1)$, with $T_o$=100K.  Owing to the weakness of the magnetic
signal ( $\le$ 10 \% of the nuclear one), one also becomes
sensitive to multiple scattering effects. This is the main reason
why the background in the difference signal remains weakly
negative even at high energy, where the detailed balance factor
has a negligible effect. Difference scans performed far away from
the antiferromagnetic wave vector at {\bf Q}=(0.275,0.275,14)
(Fig.~\ref{Fig-Escan}.a) and {\bf Q}=(0.5,1.8,4.7)
(Fig.~\ref{Fig-Escan}.b) show the typical energy dependence of the
negative background. On top of the above described negative
background, the enhanced magnetic intensity appears in energy
scans at the AF wave vector at {\bf Q}=(0.5,0.5,14)
(Fig.~\ref{Fig-Escan}.a) and {\bf Q}=(0.5,1.5,4.7)
(Fig.~\ref{Fig-Escan}.b). The resonant magnetic signal is located
at $E_r$=42 meV and exhibits a Gaussian profile with an energy
width of $\sigma_r \sim$ 13 meV (FWHM).
From constant-Q scans (Fig.~\ref{Fig-Escan}), it appears that the
magnetic intensity at all wave vectors is limited at low energy by
a {\it spin-gap} of $\sim$ 32 meV. Below the spin gap, the
magnetic signal is, at least, less than 1/4 of its value at
$E_r$=42 meV. This value agrees with reports on YBCO for similar
doping levels \cite{bourges96,science_Bourges00,reznik}.


\begin{table}[t]
\begin{tabular}{c|cccc}
\hline
Refs.&T$_c$(K)&E$_r$(meV)& $\sigma_r$(meV)& $\sigma$(meV)\\ 
\hline
Fong {\it et al}\cite{Nature_Fong99}&91&43&13 $\pm$ 2&12 $\pm$ 2\\
Present study &87&42&13 $\pm$ 2&11 $\pm$ 2\\
He {\it et al}\cite{he}& 83&38&12 $\pm$ 2&10 $\pm$ 2\\
Capogna {\it et al}\cite{Lucia}&70&34&8 $\pm$ 1&5 $\pm$ 1\\
\hline
\end{tabular}
\caption{ Characteristic energy and energy width $\sigma_r$ (FWHM: Full Width at Half maximum) of the magnetic resonance peak reported until now in Bi2212 samples. $\sigma$ stands for the intrinsic FWHM of the resonance peak, after deconvolution by the energy resolution $\sigma_{\omega}\simeq 6$ meV. }
\label{Tab-Res}
\end{table}

\begin{figure} [t]
\centerline{\includegraphics[width=8.5 cm]{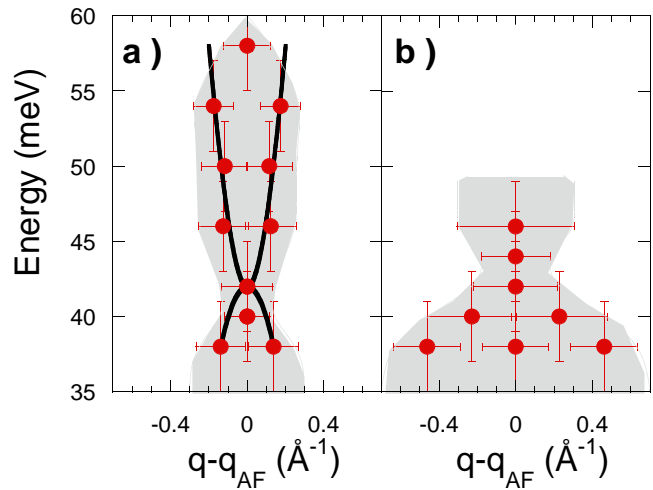}}
\caption{Dispersion of the resonant magnetic excitations deduced from constant energy scans: 
a) along the (130) direction (the full lines are described in the text), b) along the (110) direction. 
Horizontal and vertical error bars 
stand for the half width at half maximum of the signal and the energy resolution, respectively.
}
\label{Fig-dispersion}
\end{figure}

In addition to the constant Q-scans shown in Fig.
~\ref{Fig-Escan}, systematic constant-energy scans have been
performed in each scattering plane (Fig.~\ref{Fig-Qscan}) to
characterize the q-dependence of the magnetic intensity. Along the
(130) direction, the magnetic resonance peak at 42 meV  is
centered at the AF wave vector and displays a Gaussian lineshape
on top of a sloping background (Fig.~\ref{Fig-Qscan}.b). A similar
signal can be observed at 40 meV, whereas at slightly lower
energy, 38 meV, the magnetic response weakens and starts to
broaden. Although the spin response is not clearly peaked at an
incommensurate wave vector, the lineshape of the signal at this
energy can be fitted to a double-peak structure.  On decreasing the
energy transfer further, no sizeable magnetic response can be
detected at 32 meV. Nonetheless, within the error bars one cannot
exclude a weak magnetic response of magnitude at most half of the
one measured at 38 meV. It is worth noting that below 36 meV, the
large nuclear background makes the detection of a magnetic
response (if any) particularly difficult. The broadening of the
magnetic response is not observed exclusively  below the energy of
the magnetic resonance peak. At higher energy  (scans at 46 and 50
meV in Fig.~\ref{Fig-Qscan}.b), the magnetic response also
broadens, but in contrast to the measurements  at 38 meV, the
magnitude of the signal remains comparable with that of the magnetic
resonance. While at 54 meV a magnetic response with a double peak
profile can be observed, at slightly higher energy, 58 meV, the
magnetic response seems to be weak and commensurate, suggesting
that the spin excitations  may move back to the AF wave vector.
The broadening of the magnetic response above and below $E_r$=42
meV is consistent the differential energy scans performed away
from the AF wave vector at {\bf Q}=(0.52,1.56,4.7) and {\bf
Q}=(0.54,1.62,4.7) (Fig.~\ref{Fig-Escan}.b). Indeed, at these wave
vectors, the magnetic scattering is widely spread in energy, but
shows a slight dip at 42 meV. The combination of energy scans and
constant-energy scans along the (130) direction is consistent with
the existence of resonant spin excitations dispersing upward and
downward, as previously observed in the YBCO system. Guided by the
data in YBCO \cite{science_Bourges00,PRL_Pailhes04}, we then model
both dispersions of Fig. \ref{Fig-dispersion}.a by the relations:
$E_\pm = \sqrt{(E_r)^2 \pm (\alpha_\pm (q-q_{AF}))^2}$ where (-)
and (+) correspond to the downward and upward dispersions,
respectively. One obtains $\alpha_- = 130$  meV.\AA\ and $\alpha_+
= 200$  meV.\AA\ in good agreement with the dispersions in YBCO
obtained for similar doping levels
\cite{science_Bourges00,PRL_Pailhes04,PRL_Reznik04}.

Along the (110) direction, the evolution of the momentum
distribution of the magnetic signal as a function of the energy
exhibits the same trends as those observed along the (130)
direction, with a broadening at both high and low energies (46 meV
and 40 meV in Fig.~\ref{Fig-Qscan}.a). A slight double-maximum structure
also shows up in the energy scan at {\bf Q}=(0.425,0.425,14)
(Fig.~\ref{Fig-Escan}.a). But a closer inspection of the data
reveals marked differences. First,  the signal begins to broaden
already at 40 meV, and becomes almost twice as broad at 38 meV. At
this energy, the lineshape of the magnetic response is now rather
different, with a more complex structure that can be qualitatively
described in terms of a central peak at the AF wave vector
surrounded by two satellites of similar magnitude at {\bf Q}=$\rm
(-0.5 \pm \delta,-0.5 \pm \delta,14)$ with $\delta$=0.2.
As a result of this expansion, a magnetic
signal can still be observed in differential energy scans at {\bf
Q}=(0.35,0.35,14) (Fig.~\ref{Fig-Escan}.a), i.e at a large
distance in reciprocal space from the AF wave vector.

\begin{figure} [t]
\centerline{\includegraphics[angle=0,width=8.5 cm]{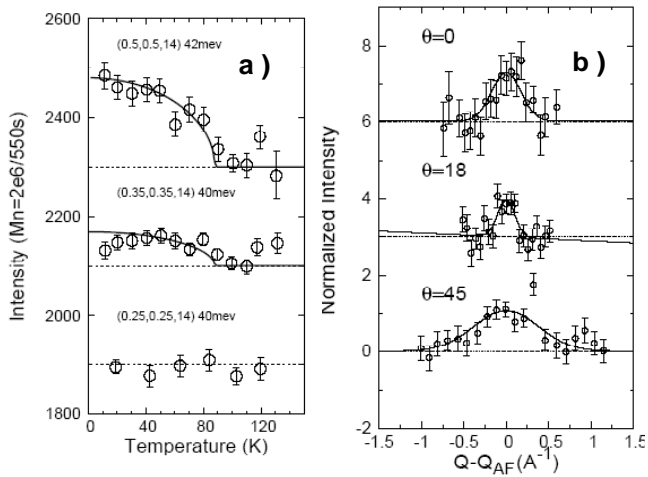}}
\caption{a) Temperature dependencies at different wave vectors
and energies. b) Differences between constant energy scans at 10 K and 100K. Scans are
performed at 40 meV around the AF wave vector along 3 different directions: (100), (130),(110).
The label $\Theta$ corresponds to the angle between the scanning direction of the (100)
direction (see Fig. \ref{continuum}.f). Data are normalized so that the intensities at
the AF wave vector are the same.}
\label{Fig-40meV}
\end{figure}

As emphasized above, one of the hallmarks of the resonant spin
excitations in SC cuprates is their peculiar temperature
dependence \cite{Sidis_review}. In this respect, the behavior of
Bi2212 is very similar to the phenomenology reported in YBCO. The
magnetic resonance peak at 42 meV disappears steeply at $T_c$, as
can be seen in Fig.~\ref{Fig-40meV}.a at 42 meV. The temperature
dependence measured at {\bf Q}=(0.35,0.35,14) and 40 meV suggests
a similar change at $T_c$, indicating that the differential signal
at this wave vector is also of magnetic resonant type. At the
background position, {\bf Q}=(0.25,0.25,14), the T-dependence does
not show any indication of a magnetic signal.

We now discuss some general aspects of the spin excitations of
Bi2212. While the overall layout of the magnetic spectrum of
Bi2212 is quite similar to that of YBCO, well-defined
incommensurate magnetic peaks were not observed in Bi2212. Several
reasons can be put forward to explain this difference. Obviously,
the weak signal-to-noise ratio of the measurements in Bi2212
limits the possibility to see details of the q-dependence. In
particular, we had to work with a broad q-resolution in order to
pick up enough intensity. (Attempts with improved q-resolution
were not successful because of insufficient intensity.) However,
scans with similar resolution and scattering plane
\cite{science_Bourges00} did reveal incommensurate peaks below
$E_r$ in YBCO. Therefore, the origin of the difference between
both systems is not purely instrumental, but at least in part
intrinsic to Bi2212. It was previously noticed\cite{Sidis_review} 
that energy scans in Bi2212 exhibit a width larger than the energy resolution for
all doping levels, as shown in Table \ref{Tab-Res}, whereas the
energy width of the odd resonance peak in YBCO is
resolution-limited at optimal doping
\cite{PRL_Pailhes04,PRL_Reznik04}. The intrinsic width of the mode
blurs the details of the mode dispersion in Bi2212. If we
nevertheless try to fit the data away from $E_r$ with a double
peak structure, one obtains dispersive excitations both below and
above $E_r$ as shown in Fig.~\ref{Fig-dispersion}, as previously
reported in YBCO
\cite{science_Bourges00,PRL_Pailhes04,PRL_Reznik04}. 
The overall energy and momentum
dependences of resonant spin excitations then exhibit the typical
hour-glass lineshape characterizing the resonant mode dispersion
in YBCO, as evidenced by the color maps of Figs.
\ref{continuum}.d-e deduced from our data along the two
q-directions of Fig. \ref{Fig-Qscan}. At $E_r=$42 meV, the
magnetic signal shrinks around $\rm {\bf q}_{AF}$. Further, below
$E_r$, additional magnetic excitations appear to develop in a
momentum region far from $\rm {\bf q}_{AF}$. As shown by color
maps of Figs. \ref{continuum}.d-e, the spectral weight of these
excitations is maximum along the diagonal direction, leading to an
anisotropy that we discuss in section \ref{aniso}.

\begin{figure}[t]
\centerline{\includegraphics[angle=0,width=7 cm]{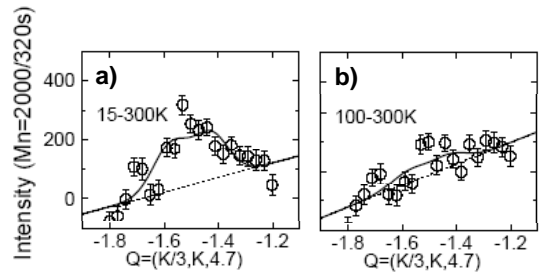}}
\caption{Differences between constant energy scans at 46 meV
performed between : a) 10 K and 300K, b) 100 K and 300K.
Measurements were carried out on the spectrometer IN8. The solid
line corresponds to the fit of the data by a double Gaussian
functional form, on top of a sloping background (dashed line). The
background at the AF wave vector is set to 0 to get rid of the the
variation of the thermal enhancement of the nuclear background and
to allow a direct comparison of both results.}
\label{Fig-QscanTdep}
\end{figure}

As pointed out at the beginning of this section, we focus on the
enhancement of the magnetic response in the SC state. The main
reason  is that previous INS measurements in optimally doped
Bi2212 \cite{Nature_Fong99} show that normal state excitations are
not measurable around 40 meV. These early measurements are in
agreement with the results in YBCO for the same doping level
\cite{bourges96,fong96}. In YBCO \cite{miami}, a systematic study
as a function of hole doping indicates that the normal-state AF
fluctuations weaken continuously with increasing doping level and
fall below the detection limit above optimal doping \cite{miami}.
Note that the fact that the magnetic fluctuations are not sizeable
above $T_c$ does not necessarily mean that they are absent, but it
rather suggests that they are much weaker and/or broader in the
normal state than in the SC state. For instance in weakly
overdoped YBCO, the typical magnitude of the spin fluctuations
left in the normal state at the resonance energy is estimated to
be one order of magnitude weaker that magnetic resonance peak in
the SC state \cite{bourges96}.
In our slightly overdoped Bi2212 sample, the upper limit on the
magnitude of the magnetic signal left in the normal state
extracted from our data is about 1/4 of the signal in the SC state
at 46 meV (Fig. ~\ref{Fig-QscanTdep}). This is consistent with a
previous report at 43 meV in an optimally doped Bi2212 sample
\cite{Nature_Fong99}.

\section{\label{fingerprint}Fingerprints of the electron-hole spin flip continuum}

\begin{figure} [t]
\centerline{\includegraphics[angle=0,width=9 cm]{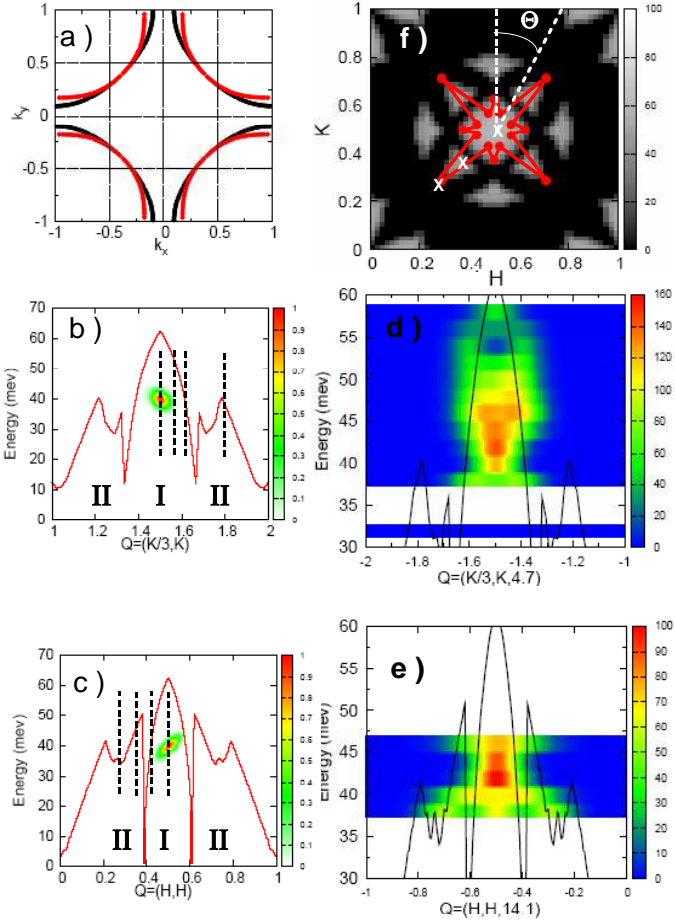}}
\caption{ (color online) a) Fermi surfaces in nearly optimally doped Bi2212 deduced from ARPES
measurements \cite{PRB_Kordyuk03, Mesot_PRL99}. (red=bonding band, black=anti-bounding band).
Threshold of the electron-hole spin flip continuum in the SC state along different directions:
b) (130), c) (110). Below the threshold of the continuum, one distinguishes two distinct areas
labelled respectively I and II as in \cite{PRL_Pailhes04} (see text). The vertical dashed lines indicate
the location of the energy scans which were performed. The projection of the resolution ellipsoid
at 40 meV is also represented. d-e) Color maps showing the enhanced magnetic response in the
SC state as deduced from the fit of the constant energy scan, in addition to the threshold of
the continuum. f) Area occupied by the continuum at 40 meV (black).
The red points indicate the extension of the magnetic as deduced from Fig.~\ref{Fig-40meV}.b
(see text) and the white crosses show where the temperature dependencies of the scattering intensity were measured.}
\label{continuum}
\end{figure}

In order to elucidate the relationship between the resonant spin
excitations in the SC state and the gapped Stoner continuum, we
have computed the threshold of the continuum in the odd channel
for the two main directions along which most of the INS
measurements were carried out (see Fig.~\ref{continuum}.b-c):
(130) and (110). One has to pay attention to the bilayer structure
that affects both spin and charge properties. The motion of
electron between the two layers in a bilayer unit leads to the
formation of anti-bonding (a) and bonding (b) states, and
consequently to a splitting of the Fermi surface, as shown in
Fig.~\ref{continuum}.a. The odd neutron scattering channel
originates from electronic interband spin flip excitations, and
the threshold of the Stoner continuum in the SC state is defined
as the minimum $[ E^a_{k}+E^b_{k+q}]$, where
$E^{a,b}_k=\sqrt{\xi^{a,b 2}_k+\Delta_k^2}$ is the electronic
dispersion relation in the SC state. $\Delta_k=\Delta_m(\cos
k_x-\cos k_y)/2$ stands for the $d$-wave SC gap  and the bare
electronic dispersion relation is described using a simplified
tight-binding expression: $\xi^{a,b}_k=2t (\cos k_x+\cos k_y) -4t'
\cos k_x \cos k_y + 2t"(\cos 2k_x-\cos 2k_y)\pm \frac{1}{4}
t_{\perp} (\cos k_x-\cos k_y)^2 -\mu $. For a nearly optimally
doped Bi2212 sample with a $T_c$ of 87 K, we used the following
parameters determined by ARPES: $\Delta_m$=35 meV
\cite{Mesot_PRL99} and
$\{0.219,0.108,0.207,-0.95 \}$ for the parameters $\{
t',t",t_{\perp},\mu \}$ in units of $t$ \cite{PRB_Kordyuk03}. The
chemical potential corresponds to a hole doping level determined
from the SC transition temperature according to the
phenomenological relationship of Tallon {\it et al}
\cite{PRB_Tallon03}.

Before proceeding to the comparison of the locations of the gap in
the Stoner continuum and the observed resonant spin excitations,
one needs to mention that there are two ways to determine the
absolute scale of electronic parameters. The value of these
parameters in absolute units (meV) strongly relies on the estimate
of the value of the nearest-neighbor hopping parameter $t$, which
in turn can be extracted from the Fermi velocity $v_F$ measured in
ARPES experiments. Along the nodal direction, $v_F\simeq$ 2.0
eV\AA \cite{arpes} is related to the {\it bare} Fermi velocity by
$v_F=v^{\circ}_F/(1+\lambda)$ where $\lambda$ describes the
electronic interactions related to the real part of the electronic
self-energy \cite{Eschrig_review,arpes}. In principle,
$v^{\circ}_F$ can be determined by the band structure given by the
Local Density Approximation (LDA). In a self-consistent
experimental procedure using the Kramers-Kronig transformation to
determine the self-energy \cite{PRB_Kordyuk03}, the energy scale
$t$ is deduced from an estimate of the {\it bare} Fermi velocity
along the nodal direction, $v^{\circ}_F\sim$ 4.0 eV.\AA
\cite{PRB_Kordyuk03}, yielding $t$=397 meV,
in good agreement with LDA predictions. An alternative approach to
analyse the ARPES data essentially defines an effective band
structure describing only low energy electronic excitations (see
{\it e.g.} \cite{chang}). The hopping parameter determined in this
way is about a factor of two smaller: $t \sim 200$ meV. This does
not significantly change the lineshape of the gapped Stoner
continuum and can be ignored in this section. However, it is
important for a quantitative description of the S=1 mode origin
(see section \ref{section-RPA}).

Figures~\ref{continuum}d-e show color maps of the magnetic
intensity deduced from fits to the constant energy scans of
Fig.~\ref{Fig-Qscan}. The threshold of the gapped Stoner continuum
is superimposed on the magnetic signal. For the sake of
simplicity, one can define two distinct areas below the continuum
(Fig.~\ref{continuum}.b-c). In area I, the threshold of the
continuum decreases away from the AF wave vector, exhibiting a
dome-like shape. Area II corresponds to the remaining part of the
gapped portion of the phase space. Along the (130) direction, the
magnetic response in the SC state is confined to area I and
appears to asymptotically approach the threshold of the continuum.
The signal vanishes below the detection limit in areas where
continuum excitations are possible according to the calculation.

Along the (110) direction, the situation is more complex. As shown
in Fig.~\ref{continuum}c,e, area I is smaller than along (130), so
that its spread in momentum becomes similar to the q-width of the
experimental resolution ellipsoid (Fig.~\ref{continuum}.c).
Moreover, the intrinsic energy width discussed above scrambles the
observed magnetic response, limiting an accurate determination of
the exact number of branches of the magnetic dispersion and their
precise location. Some interesting observations can be made
nonetheless. In particular, along (110) area II extends up to
energies in the range 35-48 meV, higher than along (130). The
extra magnetic signal along the (110) at 38 meV may hence be
attributable to secondary spin excitations in area II, in addition
to those observed in area I. In slightly underdoped YBCO$_{6.85}$,
such a secondary magnetic contribution has been observed in area
II, albeit at higher energies around $\sim$ 54 meV
\cite{PRL_Pailhes04}. There it was also shown that the continuum
gives rise to almost vertical {\it silent bands}
\cite{PRL_Pailhes04}, where the intensity of collective magnetic
modes is suddenly suppressed presumably due to the decay into
elementary electron-hole spin flip excitations. This scenario was
further confirmed theoretically within the spin exciton model
\cite{PRL_Eremin05}. Interestingly in Bi2212 the scan along (110)
at 38 meV (Fig.~\ref{Fig-Qscan}.a) displays a slight minima at the
planar wave vectors {\bf q}$\simeq$(-0.4,-0.4) and {\bf
q}$\simeq$(-0.6,-0.6), which could be indicative of silent bands
(see Fig.~\ref{continuum}.c) in a location that coincides with the
one expected based on the Fermi surface topology
\cite{PRB_Kordyuk03}.

The comparison between the location of the observed magnetic
excitation spectra and the momentum shape of the Stoner continuum
of the $d$-wave superconductor thus shows that spin excitations
are only present in the gapped regions of the continuum. We
emphasize that the locus of the Stoner continuum was computed from
the electronic excitations directly measured by ARPES for the same
doping level \cite{PRB_Kordyuk03}, without adjustable parameters.
This interpretation is also qualitatively consistent with the
converse computation of the imaginary part of the dynamical
susceptibility from ARPES data in the framework of the spin
exciton scenario \cite{condmat_Campuzzano06,borisenko}. These
computations indeed indicate the existence of two magnetic
resonant modes, located in area I and area II respectively. The
detailed momentum shape of the magnetic spectra (Figs.
\ref{continuum}.d,e) further suggests that the resonant spin
excitation spectra exhibit a hourglass lineshape mainly confined
to area I.
Overall, these results are in good agreement with previous studies
carried out in YBCO system
\cite{PRL_Pailhes04,stock,hayden04,hinkov06}, although the
intrinsic energy width of the magnetic modes in Bi2212 obscures
some of the features.

\section{\label{aniso}Anisotropy of the magnetic response}

Another interesting experimental feature is the in-plane
anisotropy of the magnetic response. Figure~\ref{Fig-40meV}.b
shows scans performed at 40 meV around the AF wave vector along 3
different directions: (100), (130), and (110). The data are
labelled as a function of the angle $\Theta$ between the direction
of the scans and the (100) direction (see Fig. \ref{continuum}.f).
Furthermore, the data are plotted as a function of the reduced
distance to the AF wave vector (in units of \AA$^{-1}$), so that
the q-widths of the signals are directly comparable. In all
directions, we fit the magnetic peak to a single Gaussian. From
$\Theta=0^{\circ}$ where $\Delta_q=0.4$~\AA$^{-1}$ (Full Width at
Half Maximum) to $\Theta=18^{\circ}$ where
$\Delta_q=0.23$~\AA$^{-1}$, the magnetic signal exhibits a
Gaussian profile, with a reduction of its q-width at
$\Theta=18^{\circ}$. On the contrary at $\Theta=45^{\circ}$, the
q-width of the signal more than doubles,
$\Delta_q=0.9$~\AA$^{-1}$, indicating a net anisotropy of the
magnetic response along the (110) direction. Along that direction
and at the resonance energy 42 meV, one obtains $\simeq$
0.45~\AA$^{-1}$ for the intrinsic q-width of the magnetic signal
after deconvolution from the resolution function in agreement with
previous reports \cite{Nature_Fong99,he}.

Concerning the origin of the observed anisotropy of the magnetic
response along the diagonals, one needs to keep in mind that the
structure of Bi2212 is not simply tetragonal. Indeed, Bi2212 is
actually an orthorhombic system with an incommensurate distortion
likely of composite type \cite{etrillard}, because the lattice
parameters of the CuO$_2$ planes and those of the BiO$_2$ planes
do not match. The orthorhombic axes are along the diagonals of the
square lattice: the sample growth direction, {\it i.e.} $a^*_{ortho}$, 
was kept perpendicular to the scattering plane. The direction
(110) in the tetragonal notation we have adopted in this article
correspond to $b^*_{ortho}$. The incommensurate modulation is
given by $0.21 b^*_{ortho} + c^*$. Because of this modulation, the
Bi2212 samples are not twinned, although the in-plane lattice
parameters are not very different. It is therefore conceivable
that the observed magnetic anisotropy along $b^*_{ortho}$ is
related to this 1D structural anisotropy. It is interesting to
remark here that in the YBCO system, the magnetic response also
exhibits a 1D-like anisotropy of the magnetic intensity, which is
maximal along a particular in-plane direction , the (100)
direction perpendicular to the CuO chains \cite{Nature_Hinkov04}.
As the CuO chains in YBCO, the BiO$_2$ planes in the Bi2212 system
play the role of charge reservoir. We therefore cannot exclude
that the diagonal anisotropy is a feedback of the structural
distortion on the magnetic properties. However, there are
differences in the magnetic anisotropy of both systems. In the
YBCO system, the anisotropy near optimal doping is mostly related
to the intensity \cite{Nature_Hinkov04}. Here, the q-extension of
the magnetic signal appears to be different along the two
directions. Further INS experiments
in which the two in-plane directions $a^*_{ortho}$ and
$b^*_{ortho}$ are studied under the same resolution conditions,
following prior work in YBCO \cite{Nature_Hinkov04}, will be
needed to shed light on the origin of the anisotropy of the
magnetic response in Bi2212.

We report in Fig.~\ref{continuum}.f the momentum dependence of the
threshold of the continuum at 40 meV. The area I centered at the
AF wave vector exhibits a diamond shape, whereas the area II
splits into 8 segments (4 along the diagonals and 2 along
$a^*$ and $b^*$). The red points in the figure indicate the total
momentum expansion of the magnetic responses (roughly twice the
full width of the signal at half maximum), as deduced from
Fig.~\ref{Fig-40meV}.b. The data have been symmetrized to account
for the fourfold symmetry of the CuO$_2$ plane although, as we
have remarked above, it is not known whether the magnetic signal
respects the square lattice symmetry. Anyway, one notices in
Fig.~\ref{continuum}.f that the magnetic signal matches the area I
along (100) but extends into area II along the diagonals. However,
the intensity of the magnetic signal in area II along the (110)
direction is of the same order as the error bars in the
measurement performed along the (100) direction. We therefore
cannot rule out the existence of a weak response in area II also
along the (100) direction, but this extra magnetic response (if
any) vanishes along the (130) direction where area II is absent.


Finally, whatever the origin of the anisotropy, its observation
allows us to solve an old puzzling issue. In the initial work of
Fong {\it et al.} in optimally doped Bi2212 \cite{Nature_Fong99},
it was found that the energy-integrated spectral weight at $\rm
{\bf q}_{AF}$ of the magnetic resonance peak was similar in
optimally doped Bi2212 and YBCO ($\sim 2 \mu_B^2 / f.u$), whereas
the local spin susceptibility (integrated in momentum space) was
actually four times larger in Bi2212 than YBCO. This discrepancy
came from the estimation of local spin susceptibility based on the
assumption that the momentum distribution of the magnetic response
was isotropic in Bi2212, as it is the case at $\rm {\bf q}_{AF}$
in YBCO. Since the measurement of the momentum width of the
magnetic signal along the (110) direction of Bi2212 was found to
be twice as large as in YBCO, the local susceptibility was
estimated to be four times larger. Considering
Fig.~\ref{continuum}.f, the momentum shape at the resonance energy
found in  optimally doped YBCO typically corresponds to only area
I, and extra scattering in area II occurs at larger energy
\cite{PRL_Pailhes04}. In optimally doped Bi2212, this extra
magnetic scattering in area II is found around the same energy as
the resonance peak, yielding a much broader peak in q-space.
Therefore, the present study shows that the assumption of an
isotropic magnetic response was not correct, leading to an
overestimation of the local spin susceptibility. This implies that
the q-integrated spectral weight is likely quite similar in both
systems.


\section{\label{section-RPA}Origin of the $S=1$ collective mode: RPA description}

Based on the energy and momentum distribution of the resonant spin
excitations in the SC state, we have shown above that our INS data
exhibit fingerprints of the $d$-wave gapped Stoner continuum. This
puts constraints on the origin of the $S=1$ collective mode.
However, as far as the electron-hole spin flip continuum is taken
into account, both itinerant and localized spin models can, in
principle, describe the data. Since the charge excitation
spectrum and the SC gap are well-known in the Bi2212 family from a
considerable amount of ARPES data, this system is the right
candidate to test these scenarios quantitatively. Here, we apply
the spin-exciton model
\cite{Eschrig_review,PRL_Chubukov99,norman,PRB_Onufrieva02,PRB_Schnyder04,PRL_Eremin05}
using the measured band structure parameters and SC gap.
Furthermore,  within that framework, one can relate the energy
width of the resonance peak to the spatial distribution of the SC
gap as reported by Scanning Tunnelling Microscopy (STM)
\cite{PRL_McElroy05,Nature_Davis06}. We then investigate the
magnetic excitation dispersion along the two directions studied by
INS in order to put this comparison on a quantitative footing.

In the spin-exciton scenario, the spin susceptibility takes an RPA-like form \cite{Eschrig_review} :
\begin{equation}
\chi({\bf q},\omega) =  \frac{\chi ^0({\bf q},\omega)} {1- V_{{\bf q}}\chi ^0({\bf q},\omega)}
\label{RPA}
\end{equation}

$\chi^0({\bf q},\omega)$ stands for the standard non-interacting
spin susceptibility of a superconductor
\cite{Eschrig_review,PRL_Chubukov99,norman,PRB_Onufrieva02,PRB_Schnyder04,PRL_Eremin05}.
The interaction $V_{\bf q}$ that enhances the magnetic response
can be interpreted as: (i) the on-site Coulomb repulsion on copper
$U$ in  weak coupling models ($U \le 8t$, $8t$ being of the order
of the band width), (ii) the AF superexchange coupling $-J({\bf
q})=-2 J(\cos q_x+\cos q_y)$ in strongly correlated models ($U \ge
8t$, such as in the t-t'-J model, with $J=4t^2/U \sim$ 120 meV)
\cite{norman,PRB_Onufrieva02}, (iii) an effective spin-fermion
coupling $g({\bf q})$ in phenomenological models
\cite{PRL_Chubukov99,PRL_Eremin05}.

As before, we limit our calculations to the odd excitations. In
this channel $\chi ^0({\bf q},\omega)$  involves two-particle
excitations between bonding (b) and anti-bonding (a) states:
$\chi^0=(\chi^0_{ab}+ \chi^0_{ba})/2$. To calculate $\chi^0({\bf
q},\omega)$, we use a SC gap with $d_{x^2+y^2}$ symmetry and a
normal-state tight binding dispersion previously defined in
section \ref{fingerprint}. In order to facilitate the numerical
calculations, we used a small damping parameter of 2 meV, which is
smaller than the energy width of the INS resolution. Finally, we
use the following phenomenological form for the interaction
$V_{{\bf q}}=U_{eff}-2J_{eff}(\cos q_x+\cos
q_y)$\cite{PRL_Eremin05}. This allows us to capture the different
physical origins of the interaction.

For any given wave vector ${\bf q}$, a sharp magnetic excitation
shows up at an energy $\Omega_r$ when the conditions $1-V_{{\bf
q}}Re \chi^0({\bf q},\Omega_r)=0$ and $Im \chi^0({\bf
q},\Omega_r)=0$ are fulfilled.
For a given magnitude of the SC gap amplitude $\Delta_m$, the
measured energy position of the odd magnetic resonance peak at
$\rm {\bf q}_{AF}$=$(\pi,\pi)$ constrains the sum of the
interaction parameters $V_{{\bf q}_{AF}}=U_{eff}+4J_{eff}$,
whereas the shape of the dispersion is controlled by their ratio
$J_{eff}/U_{eff}$.

\begin{figure} [t]
\centerline{\includegraphics[angle=0,width=6 cm,]{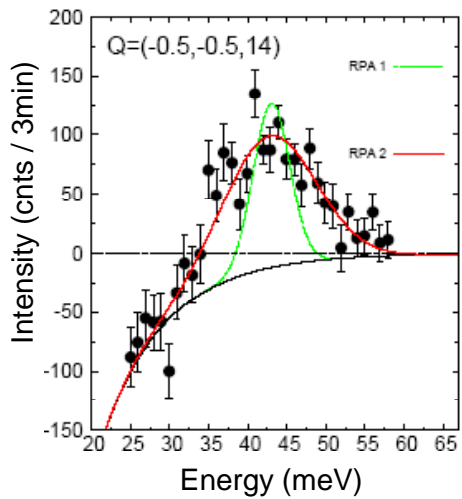}}
\caption{ Resonant magnetic modes in nearly optimally doped $\rm
Bi_2 Sr_2 Ca Cu_2 O_{8+\delta}$ ($T_C$=87 K) measured at ${\bf Q}
=(0.5,0.5,14)$ (as in Fig. \ref{Fig-Escan}.a). The solid lines
correspond to computed imaginary part of the susceptibility
convoluted with the energy resolution function: the green line was
obtained using a a single $d$-wave SC gap (RPA 1) and the red one
with a distribution of $d$-wave SC gaps (RPA 2). In the latter
case, the SC gap average is $\Delta_{m}$=35 meV with a full width
at maximum of 14 meV.} \label{energy}
\end{figure}

Let us start by studying the odd magnetic resonance peak at the
planar AF wave vector $\rm {\bf q}_{AF}$. In Fig. \ref{energy}, we
compare the measured resonant peak energy scan at $\rm {\bf
q}_{AF}$ and the spin excitation spectrum computed with
Eq.~\ref{RPA} and convoluted with the  energy resolution of the
spectrometer. For a slightly overdoped sample with T$_c$=87K,
$\Delta_m$ is set to 35 meV, yielding a threshold of the Stoner
continuum of $\omega_c$=62 meV at $\rm {\bf q}_{AF}$. To obtain a
resonance energy at the proper energy, $\Omega_r$=42 meV, well
below $\omega_c$, one needs to adjust the interaction $V_{{\bf
q}_{AF}}$=1070 meV, a value significantly exceeding the hopping
parameter $t$=397 meV. A systematic investigation yields a ratio
$V_{{\bf q}_{AF}}/t=2.7$, independent of $t$ in the range 200-400
meV, in agreement with Ref. \onlinecite{PRL_Eremin05}.

At $\rm {\bf q}_{AF}$, the computed spin excitation spectrum is
dominated by the spin exciton, whose contribution to the imaginary
part of the dynamical spin susceptibility reads: $Im \chi( {\bf
q}_{AF},\omega)= W_r \delta(\omega-\Omega_r)$ for $\omega > 0$.
The measured magnetic resonance peak should be given by the
convolution product of this $\delta$-function by the Gaussian
resolution function of the instrument: it should take the form,
$Im \chi({\bf q}_{AF}, \omega)= W_r \exp(-4\ln 2
\frac{(\omega-\Omega_r)^2}{ \sigma _\omega^2})$ where
$\sigma_\omega \simeq 6$ meV is the energy resolution (as shown by
the green line (RPA 1) in Fig. \ref{energy}). Clearly, the energy
width of the resonant mode is significantly broader than the
expected spectrum, showing that there is an additional
contribution to the broadening. This discrepancy is not specific
to our sample. Table \ref{Tab-Res} shows the corresponding energy
position and energy width of all Bi2212 samples investigated so
far by neutron scattering. For all doping levels, the measured
resonant mode is broader than the calculated resolution-limited
RPA peak. For a long time now, it has been argued
\cite{Sidis_review} that this broadening might be related to the
electronic inhomogeneity of the Bi2212 system, which also
manifests itself in the SC gap distribution observed in the real
space by STM \cite{PRB_Howald,PRL_McElroy05,Nature_Davis06}.
Interestingly, when going into the overdoped regime,  the
broadening of the INS magnetic resonance peak \cite{Lucia} and the
SC gap distribution \cite{Nature_Davis06} are reduced
simultaneously (see Table \ref{Tab-Res} for the resonance peak).

\begin{figure} [t]
\centerline{\includegraphics[angle=270,width=6 cm,]{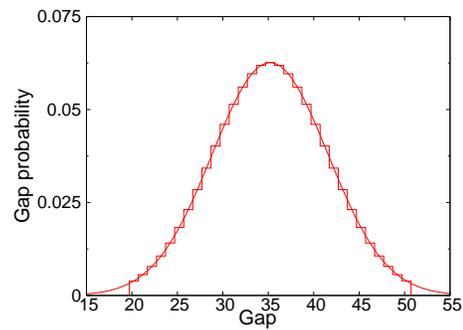}}
\caption{Spatial gap distribution considered for the calculation of $\chi({\bf q},\omega)$.
Using a Gaussian fit of this distribution, one obtains a spatial average gap of 35 meV
and a FWHM of $\sigma_{\Delta}=15$ meV corresponding to the sample with $T_c=87$ K. }
\label{histo}
\end{figure}

\begin{figure} [t]
\centerline{\includegraphics[angle=270,width=7 cm,]{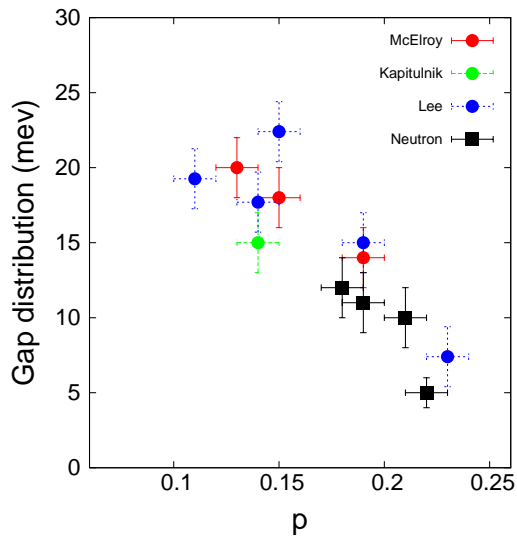}}
\caption{Width (FWHM) of the SC gap distribution measured by
STM\cite{PRB_Howald,PRL_McElroy05,Nature_Davis06} and width of the magnetic
resonance peak in INS data\cite{Nature_Fong99,he,Lucia}. }
\label{width}
\end{figure}

Based on this observation, we have tested the effect of a spatial
distribution of SC gap on the spin excitation spectrum within the
spin exciton scenario. Spatial variations of the SC gap at the
surface of Bi2212 were reported by several groups
\cite{PRB_Howald,PRL_McElroy05,Nature_Davis06}. The spatially
averaged value of the superconducting gap, $\Delta_m$, measured by
STM is consistent with other spectroscopic techniques such as
ARPES. The FWHM of the gap distribution, $\sigma_{\Delta}$,
decreases with increasing doping level (Fig. \ref{width}), from
around 15 meV at optimum doping to nearly 7 meV for an overdoped
sample with a gap average of $\Delta_m$=23 meV
\cite{Nature_Davis06}. In our calculation, we use the gap
distribution found by STM at optimal doping, and we further assume
that the interaction $V_{{\bf q}}$ is independent of the local SC
gap. Since the local gap amplitude varies, the resonance peak
position is different in different patches of the sample. We
compute the susceptibility $\chi({\bf q},\omega)$ in each patch
characterized by a different SC gap value, labelled now $\chi({\bf
q},\omega,\Delta)$. $\chi({\bf q},\omega,\Delta)$ is computed for
seventeen different SC gaps with $\Delta$ from 20 to 50 meV.
The gap distribution is implemented using the histogram shown in
Fig. \ref{histo}, which approximates a Gaussian distribution of
the SC gap with a width $\sigma_{\Delta}$.
The full susceptibility, $\chi({\bf q},\omega)$, is given by the
sum over all local susceptibilities $\chi({\bf q},\omega,\Delta)$:

\begin{equation}
\chi({\bf q},\omega) =  \int \chi({\bf q},\omega,\Delta)
\exp(-4\ln 2\frac{(\Delta-\Delta_m)^2}{ \sigma_{\Delta}^2}) d\Delta
\label{eq-RPAgap}
\end{equation}
The total spin susceptibility is  further convoluted by the Gaussian resolution function in order
to fit the data.


Using this procedure, we can simulate the broadening of the
magnetic resonance peak by adjusting the width of the Gaussian gap
distribution, $\sigma_{\Delta}$ (Eq.~\ref{eq-RPAgap}). Figure
\ref{energy} (fit: RPA 2) shows that the enhancement of the
magnetic response in the SC state at the AF wave vector can be
well accounted for by an intrinsic SC gap distribution with
$\sigma_{\Delta}$=14 meV. For an overdoped sample with $T_c$=70
K\cite{Lucia}, the same analysis gives $\sigma_{\Delta}$=5 meV. It
is striking that the gap distribution deduced from the fit of the
magnetic resonance peak matches the one reported by STM for both
doping levels (Fig. \ref{width}).
In the framework of {\it spin exciton scenario}, one can thus give
a clear explanation of the energy width of the magnetic resonance
excitation. Note that such energy broadening is absent in the YBCO
family compounds. This shows that the electronic inhomogeneity is
not generic to the cuprates. Similar conclusions about a better
homogeneity in YBCO were reached by measuring the quasi-particle
lifetimes \cite{corson} or the $^{89}$Y nuclear magnetic resonance
linewidths \cite{bobroff}.

\begin{figure} [t]
\centerline{\includegraphics[angle=0,width=8
cm,]{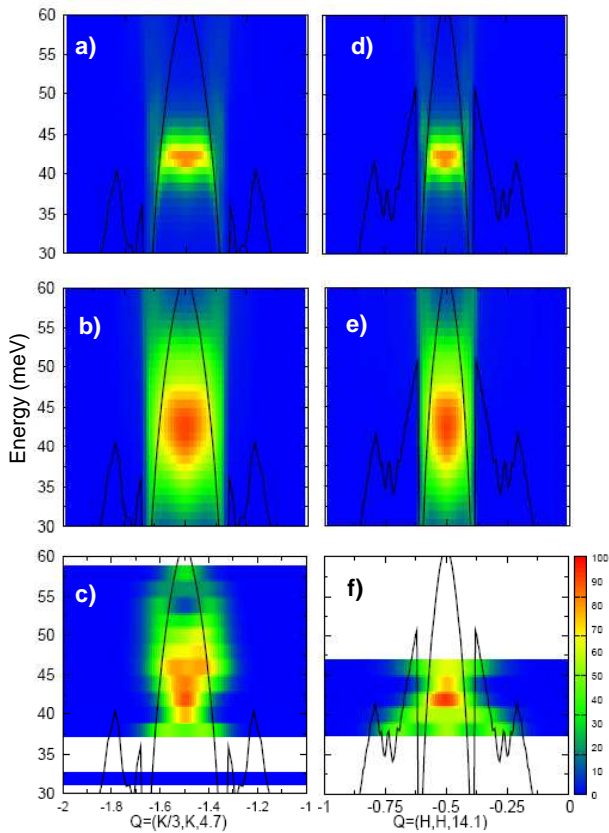}} \caption{Mappings of the imaginary part of
the dynamical spin susceptibility $Im \chi({\bf q},\omega)$: (a-c)
along the (130) direction, (d-f) along the (110) direction. $Im
\chi({\bf q},\omega)$ was computed using a spin itinerant model
(see text) assuming: (a,d) a unique SC gap for the entire system
or (b,e) the spatial distribution of SC gaps reported in Fig.
\ref{histo}. In order to enable a immediate comparison of the
computed mappings of $Im \chi({\bf q},\omega)$ and the INS
measured ones, the experimental mappings of
Fig.~\ref{continuum}.d-e are also reported (c,f). In absence of a
calibration of the INS data in absolute units, mappings are scaled
so that the maximum intensity corresponds to 100 counts.}
\label{Mapcomput}
\end{figure}

Next, we address the consequences of the SC gap distribution for
the q-dependence of magnetic excitations, still within the {\it
spin-exciton} model. We performed the calculation in the ${\bf
q}-\omega$ range covered by our experiments, {\it i.e.} along both
the (110) and (130) directions and between 30 and 60 meV. Using
the same set of parameters, we additionally adjust the ratio
$J_{eff}/U_{eff}$ to obtain the best agreement between the
experimental q-dependence (Figs. \ref{continuum}.d,e) and the
computed maps of Figs. \ref{Mapcomput}. This corresponds to the
case when $J_{eff}/U_{eff}<< 1$. We compute the maps of
Fig.\ref{Mapcomput} with $J_{eff}/U_{eff}=0.025$ for both Q
directions. As for the energy width at the AF wave vector, the
model does not describe the results in the absence of a SC gap
distribution (Fig. \ref{Mapcomput}.a and d). One obtains a rather
sharp mode dispersion downward as observed in YBCO
\cite{science_Bourges00}. In contrast (Fig. \ref{Mapcomput}.b and
e), the addition of a SC gap distribution (corresponding to the
STM data) allows us to describe the main features of the measured
magnetic excitation spectrum (Fig. \ref{Mapcomput}.c and f). For
the (130) direction, the experimental and computed mappings are
quite similar. For the (110) direction, both mappings are
consistent, even if below 42 meV the computed spectrum fails to
describe the q-broadening of the signal. We found that it is not
possible to obtain a signal around 38-40 meV in area II, whatever
the q-dependence of the interaction used.

To summarize, using the band structure and the SC gap known from
ARPES measurements \cite{PRB_Kordyuk03,Mesot_PRL99},  the
calculation of the magnetic excitations in an itinerant spin
approach can reproduce most of the  characteristic features of the
INS spectrum measured the spin excitation  spectrum in the SC
state of Bi2212. Further, the observed energy width of the
resonant spin excitations in Bi2212 is naturally explained within
this model as a result of the gap distribution reported by STM
\cite{PRB_Howald,PRL_McElroy05,Nature_Davis06}. The measured
q-dependence of the magnetic excitations can be also captured by a
broadening of dispersive excitations due to the SC gap
distribution.

Finally, we comment on a few points about the nature of the interaction, that should provide an
indication on the Hamiltonian needed to obtain a spin exciton.

First, one finds a ratio $V_{{\bf q}_{AF}}/t=2.7$ whatever the
used band structure. We consider the energy range for $t$ which is
typically 200-400 meV. Two opposite limits can then be discussed:
(i) a weak coupling approach, where the band structure is not far
from the LDA one, {\it i.e.} $t\sim 400$ meV, then the interaction
is an effective on-site Coulomb repulsion, (ii) a strong coupling
approach, where one chooses an effective band width, {\it i.e.}
$t\sim 200$ meV, reduced from the LDA value due to the strong
electronic correlations and with an interaction given by the AF
superexchange interaction $J$ (in principle measured in the AF
insulating state). In the first limit, using $t$=395 meV, the
effective interaction $V_{\bf q_{AF}}=1070$ meV can be directly
compared to the band width, $W=8t$. One finds $V_{\bf
q_{AF}}/W\approx$ 0.35, which clearly belongs to the weak coupling
approach \cite{DMFT} with an effective on-site Coulomb interaction
on copper. In the other limit, $t$ is around 200 meV and then
$V_{{\bf q}_{AF}}$=540 meV. This value can then be reduced to
$\sim 4J$ as expected in the $t-t'-J$ model
\cite{PRB_Onufrieva02}, yielding $J=135$ meV. In the insulating
state, $J$ is usually related to the on-site Coulomb interaction
as $J=4t^2/U$, yielding $U$=1185 meV. Surprisingly, the ratio
between the interaction and the band width is found to be quite
small as $U/W\approx$ 0.74, {\it i.e.} in an intermediate regime
not far from the weak coupling side.

Next, we found that $V_{\bf q}$ is rather weakly dependent on the
wave vector, since $J_{eff}/U_{eff}<< 1$. It is worth to note that
in case of a momentum independent interaction, {\it i.e.}
$J_{eff}=0$, Eq.~\ref{RPA} is unstable towards a spin-density wave
state using the band structure observed by ARPES in Bi2212. In the
other limit, still taking the same electronic parameters but only
a q-dependent interaction, {\it i.e.} $U_{eff}=0$, one observes a
resonance peak which first disperses upward and then downward when
approaching the electronic continuum delimiting the area I and II
in Fig. \ref{continuum} ({\it i.e.} the silent bands). Such an
"M"-shaped dispersion is not observed in Bi2212, although the SC
gap distribution renders the experimental situation too complex to
rule this out entirely. Further, it is shown  in Ref.
\onlinecite{norman} that the exact magnetic mode dispersion is
extremely sensitive to the detailed tight-binding parametrization
of the band structure. Therefore, depending on details of the band
structure, one can easily move from a weakly to a strongly
q-dependent interaction.

Lastly, an important issue with the spin exciton model is related
to the apparent observed "X"-shaped or hour-glass dispersion.
Within the RPA approach it is difficult to reproduce downward- and
upward-dispersing branches merging around $\Omega_r$. This is
because in the simple RPA susceptibility of Eq.~\ref{RPA}, there
is only a single pole for a given wave vector below the electronic
continuum, especially in area I. The high-energy part can be
understood by a pole condition of the RPA susceptibility below the
electronic continuum, but only in area II
\cite{PRL_Pailhes04,PRL_Eremin05}, without direct continuity with
the mode at $(\pi,\pi)$ and $\Omega_r$. This limitation of the
spin-exciton model may indicate strong-coupling effects not
capture by the RPA formalism. The capability of alternative
approximation schemes \cite{hubbard-jain,eremin} to reproduce this
feature remains to be elucidated.

\section{Concluding remarks}

We have shown that the spin excitation spectrum in the odd channel
of Bi2212 is consistent with the {\it hour glass} or "X"-shaped
dispersion previously reported in YBCO for a similar doping level
\cite{PRL_Pailhes04,PRL_Reznik04}. Most aspects of the momentum
and energy dependence can be ascribed to the existence of a $S=1$
collective mode, a {\it spin-exciton}. The study further reveals
that the imaginary part of the dynamical magnetic susceptibility
is strongly enhanced below the threshold of the electron-hole spin
flip continuum that can be derived from ARPES measurements
performed on the same system. This observation suggests that the
$S=1$ collective mode decays into elementary electron-hole spin
flip excitations when it enters the continuum. Likewise, the
influence of the gapped Stoner continuum is confirmed by a recent
study of the doping dependence of the characteristic energies and
spectral weights of the resonance peaks in both odd and even
channels \cite{Lucia}. In agreement with a similar study carried
out over a wide doping range in the (Y,Ca)BCO system
\cite{PRL_Pailhes06}, the spectral weight of both AF resonance
peaks is proportional to their reduced binding energy with respect
to the continuum. In our Bi2212 sample ($T_c$=87 K), the estimate
of the threshold of the continuum at the AF wave vector from INS
data, $\omega_c=63$ meV\cite{Lucia}, agrees remarkably well with
the value deduced from ARPES measurements.

As mentioned in the introduction, localized-spin models based on
the formation of stripe arrays of spins and charges have been also
developed to account for the spin dynamics in high-$T_c$ cuprates.
The predictions of these models can also be compared to our INS
data obtained below $T_c$ in nearly optimally doped Bi2212 sample.
However, in most of these models the effect of the
superconductivity has not been addressed theoretically
\cite{Voj04,Uhr04,Sei05}.
One stripe model that explicitly considers the presence of
superconductivity predicts only minor effects, in contrast to our
observations \cite{And}. Most of this theoretical work is based on
spin-only models, that is, they consider the magnetic response of
localized Cu spins, but typically ignore the charge degree of
freedom. However, it has been argued \cite{revue_jmt} that when
the spin-gap is large and close to the saddle-point energy, much
of the magnetic spectral weight redistributed below $T_c$ is
accumulated at the saddle point, yielding the strong commensurate
resonance peak. In principle, this argument should apply to the
Bi2212 system where the spin-gap is around 32 meV (Fig
\ref{Fig-Escan}). However, our data provide evidence that the key
factor to understand the spin dynamics in the SC state is the
momentum and energy shape of the $d$-wave electron-hole Stoner
continuum. In this picture, the spin-gap is ascribed to the energy
where the collective mode merges into the continuum. Even in a
stripe scenario, a gapped continuum delimiting the spin
excitations must also be present. Our study tells us that this
continuum has to be similar to the one obtained starting from
uniform 2D Fermi liquid theories. One more general grounds,
dual (itinerant/localized) spin models incorporating the gapped
Stoner continuum
\cite{PRB_Sega03,condmat_Prelovsek06,PRB_Onufrieva94,JETP_Eremin06}
may also be consistent with our data.

Two features of the magnetic dynamics of Bi2212 are at variance
with the YBCO system. First, the resonance peak in Bi2212
consistently exhibits an intrinsic energy width
\cite{Nature_Fong99,he,Lucia}. Within the spin-exciton model, we
were able to relate  this width to the SC gap distribution
observed by STM \cite{PRB_Howald,PRL_McElroy05,Nature_Davis06}
(Fig. \ref{width}). Second, we observed in Bi2212 an anisotropy
along the diagonal(s) (110) direction, whereas in YBCO the spin
excitation spectrum exhibits an energy dependent 1D-like
anisotropy \cite{Nature_Hinkov04} with maximum spectral weight
along $a^*$.
Whatever the origin of these anisotropies, this difference between
YBCO and Bi2212 system suggests that the specific form of the
in-plane anisotropy of the spin excitations may depend on
structural details of individual compounds.

In summary, we have determined the detailed momentum dependence of
the resonant spin excitations in the SC state of a nearly
optimally doped Bi2212 sample ($T_c$=87 K. The salient features of
this spectrum are well described in an itinerant-electron
approach. Interestingly, the intrinsic energy width and its doping
dependence are naturally explained by considering the SC gap
distribution as measured by STM. The global momentum shape of the
measured magnetic excitations is also correctly described within
the {\it spin-exciton} model in the presence of the same SC gap
distribution. The spin excitation spectrum in Bi2212 is delimited
by the gapped Stoner continuum in the $d$-wave superconducting
state, which can be directly inferred from ARPES data on the same
system. Together with prior observations in the YBCO system
\cite{PRL_Pailhes04}, this underscores the influence of the Stoner
continuum as an important ingredient in the description of the
spin dynamics in superconducting cuprates with large $T_c$.

\begin{acknowledgments}

We wish to thank A. Pautrat and C. Simon from CRISMAT laboratory (University of Caen, France)
for susceptibility measurements. We also acknowledge stimulating discussions with
S. Borisenko, A.V. Chubukov, S. Davis, I. Eremin, M. Eremin, F. Onufrieva, S. Pailh\`es,
P. Prelov\u{c}ek  and J.M. Tranquada.  This research project has been supported in part by the Deutsche
Forschungsgemeinschaft, Grant No. KE923/1–2 in the consortium FOR538, and in part
by the European Commission under the 6th Framework Programme through the Key Action:
Strengthening the European Research Area, Research Infrastructures.
Contract n$^{\circ}$: R113-CT-2003-505925.

\end{acknowledgments}


\end{document}